\documentclass[aps,preprint,superscriptaddress]{revtex4-1}
\usepackage[colorlinks=true,linkcolor=blue,urlcolor=blue,citecolor=blue]{hyperref}
\usepackage{bbding}
\usepackage{epsfig}
\usepackage{graphicx}
\usepackage{dcolumn}
\usepackage{bm}
\usepackage{ulem}
\usepackage[dvipsnames]{xcolor}
\newcommand\redsout{\bgroup\markoverwith{\textcolor{red}{\rule[0.5ex]{2pt}{0.4pt}}}\ULon}
\usepackage{amssymb}
\usepackage{amsmath}   

\newcommand{\bk}[1]{\hat{b}_{#1}}
\newcommand{\bdk}[1]{\hat{b}_{#1}^{\dagger}}

\begin{document}
%
\title{Enhancing supercurrent-based inertial sensing via interactions in atomtronic angular accelerometers}

\author{S. Carmona-López}
\affiliation{ Facultad de F\'isica, Pontificia Universidad Cat\'olica de Chile, Casilla 306, Santiago 22, Chile}

\author{A. Matos-Abiague}
\affiliation{Department of Physics and Astronomy, Wayne State University, Detroit, MI 48201, USA}

\author{F. Isaule}
\affiliation{ Facultad de F\'isica, Pontificia Universidad Cat\'olica de Chile, Casilla 306, Santiago 22, Chile}

\author{L. Morales-Molina}
\affiliation{ Facultad de F\'isica, Pontificia Universidad Cat\'olica de Chile, Casilla 306, Santiago 22, Chile}

\begin{abstract}

We theoretically investigate supercurrents of ultracold atoms in angularly ac-shaken ring lattices subjected to external rotation. Our results demonstrate how these supercurrents can be harnessed for the development of high-precision atomtronic angular accelerometers. Using both analytical and numerical approaches within the Bose-Hubbard model framework, we demonstrate that a significant net atomic current arises when the lattice driving frequency is tuned to an integer fraction of the Bloch frequency, while the current averages to nearly zero away from such a resonance. In the single-particle regime, the resonance width scales inversely with the averaging time, thereby setting a fundamental Fourier-limited bound on the measurement’s sensitivity. Strikingly, our numerical simulations demonstrate that this Fourier limit—a fundamental barrier in the non-interacting system—can be surpassed by introducing weak interactions between atoms.  In the interacting regime, the sensitivity surpasses the Fourier-limited scaling with the averaging time achieving an improvement of at least two orders of magnitude over the single‑particle scenario, and exceeding the performance of previously proposed ultracold‑atom-based angular accelerometers. 
These findings pave the way for developing new atomic-current-based inertial sensors with interaction-enhanced sensitivity.

\end{abstract}

\maketitle

\section{Introduction}

 Quantum sensors can offer significant advantages over classical devices, providing enhanced sensitivity and precision~\cite{quantumsensing, highaccuracyinertialmeasurements,salducci2024quantum, pelegri2018quantum, sabin2014phonon,morales2022quantum}. Among them, cold-atom sensors stand out for their ability to measure physical quantities with remarkable stability and accuracy, making them a cutting-edge area of research in ultracold atomic physics~\cite{quantumsensing,salducci2024quantum}.

Among cold-atom sensors,  those designed to measure inertial and gravitational signals have been the subject of intensive investigation \cite{salducci2024quantum,highaccuracyinertialmeasurements}. A particularly notable class within this category relies on the dynamic modulation of optical lattices, such as lattice shaking, in which atoms are confined. Lattice shaking has emerged as a powerful technique for manipulating cold atoms, enabling applications that range from quantum momentum-state engineering to condensate control \cite{Momentum-state-engineering}. Its implementation has led to significant advancements in the performance of atom interferometry-based sensors \cite{interferometry-shaking, weidner2018experimental,karcher2018improving,Reinforcement,reinforcement_shaken_2d_giroscopio}. Moreover, ac-driven optical lattices can give rise to resonant processes that either enhance the sensitivity of existing inertial sensors or enable new types of sensing mechanisms. A well-known example involves tilted periodic optical lattices, where gravity induces Bloch oscillations of cold atoms. In this context, tuning the driving frequency to match the Bloch frequency has been shown to significantly improve the precision of the gravity acceleration measurements \cite{blochoscillationsgravity,tarallo2012resonanttunnelingBO}.

 Extending this framework to rotational configurations, angular Bloch oscillations in ring‑shaped optical lattices have recently been investigated theoretically \cite{angularBOs}, demonstrating that angular acceleration can be accurately estimated from the angular Bloch frequency. Motivated by these results, in this work, we present and analyze a theoretical proposal for an angular accelerometer based on an angularly ac-shaken ring optical lattice. In contrast to prior studies, we propose to leverage measurements of atomic supercurrents as a promising method for detecting angular acceleration arising from external rotation. 
Specifically, when the driving frequency of the shaken lattice is a subharmonic of the angular Bloch frequency, a nonzero average atomic current emerges. Such a net transport of atoms has been observed in one-dimensional lattices under similar resonance conditions \cite{haller2010inducing-transport}. This effect provides the foundation for an {\it atomtronic} angular accelerometer, which operates by measuring supercurrents, a core phenomenon in the field of Atomtronics ~\cite{atomtronics, RMP-Atomtronics, cold-atoms-ring, persistentcurrents-review, persistent-currents-Spehner, Persistent-currents-ring-Miguzzi}.

A key advantage of our method proposed here is its reliance on the emergence of a net average current, which eliminates the need to track the spatial evolution of the atomic cloud, thereby avoiding uncertainties associated with its spreading during time evolution. However, in the non-interacting regime, the measurement sensitivity is fundamentally limited by the associated Fourier-limited bandwidth \cite{blochoscillationsgravity}, a constraint common to resonant two-level systems that ultimately bounds the precision of the angular acceleration estimate.

 To overcome this limitation, we propose leveraging atomic interactions to enhance the supercurrent's sensitivity to angular acceleration. Although typically detrimental to measurement precision in atom interferometry, such interactions can boost the sensitivity of some ultracold atomic sensors \cite{high-precision-gravity-BEC}, as demonstrated in Bose-Einstein condensates, where interatomic collisions have been shown to improve the sensitivity of cold-atom gravimeters beyond the shot-noise limit \cite{high-precision-gravity-BEC}.

In this work, we show that introducing weak atomic interactions can not only improve the sensitivity by several orders of magnitude compared to the single-particle case, but also surpass values previously predicted for angular accelerometers \cite{angularBOs}. This interaction-enhanced sensitivity establishes our atomtronic accelerometer and its potential for developing a new class of inertial sensors based on supercurrents of weakly interacting ultracold atoms.

The paper is organized as follows. Section~\ref{sec:model} introduces the theoretical framework of an angularly ac‑shaken ring lattice and formulates the atomtronic angular accelerometer. In Section~\ref{sec:method}, we present the sensing protocol based on the detection of time‑averaged supercurrents. We first analyze the single‑particle regime, where analytical and numerical calculations reveal the emergence of a finite average current at resonant drive frequencies determined by submultiples of the angular Bloch frequency, while off‑resonant responses vanish upon long-time averaging. We then extend the analysis to the non‑equilibrium dynamics following a phase quench. The section further addresses the weakly interacting regime, demonstrating that atomic interactions significantly enhance sensitivity by inducing a pronounced narrowing of the resonance. This leads to a sensitivity that surpasses the standard Fourier scaling with averaging time. We attribute this behavior to interaction‑induced dephasing and interference among Floquet modes, which is further elucidated by an effective reduced Hamiltonian description near resonance. Finally, Sec.~\ref{sec:concl} summarizes the main conclusions and outlines possible directions for future research.

\section{Theoretical Model} \label{sec:model}
\begin{figure}
    \centering
    \includegraphics[scale=0.45]{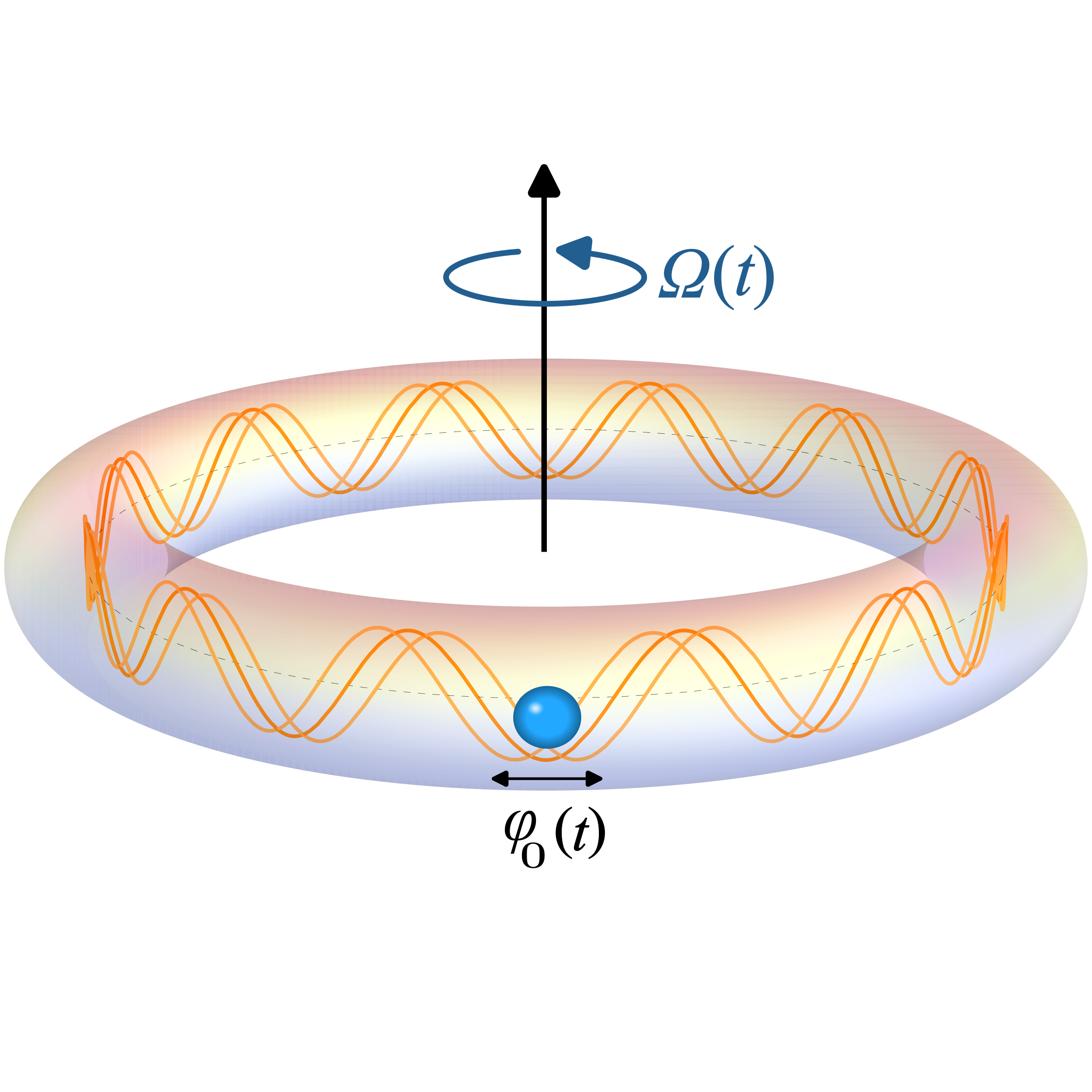}
    \caption{Schematic of the proposed setup, consisting of atoms confined in a toroidal trap effectively restricting their motion to a shaken ring optical lattice with a time-dependent angular displacement $\varphi_0(t)$. In addition, the trapped atoms experience an external rotation about the ring's symmetry axis with angular velocity $\Omega(t)$.}
    \label{fig:toroide y anillo}
\end{figure}

We consider an effective one-dimensional system of ultracold atoms in a time-periodic, angularly shaken ring lattice, subjected to an external angular acceleration. One-dimensional lattices without periodic driving can be realized by combining strong confinement in a toroidal trap with a ring lattice potential generated by two Laguerre-Gaussian beams \cite{angularBOs}. The full system (i.e., the lattice with periodic driving) is represented in Fig. \ref{fig:toroide y anillo}. 

The Hamiltonian of the single-particle system can be written as \cite{angularBOs},

\begin{equation}
        \hat{H}_0 = \frac{\hat{p}_\varphi^2}{2I_{rot}} + V\left[\hat\varphi+\varphi_0(t)\right] - \Omega(t)\hat p_\varphi\,,
    \label{hamiltoniano inicial}
\end{equation}
where $\hat\varphi$ and $\hat p_\varphi$ are the angular position and conjugate angular momentum operators, respectively. The moment of inertia is denoted by $I_{rot}$, and $\Omega(t)$ is the external time-dependent angular velocity around the ring's symmetry axis. The lattice periodic potential is represented by $V\left[\hat\varphi+\varphi_0(t)\right]$, where the time-dependent phase $\varphi_0(t)$ accounts for lattice shaking. 

 In this work, we consider a periodically shaken ring lattice, described by an angular displacement
\begin{equation}\label{Eq:phase-shaking}
\varphi_0(t) = A\cos(\omega t + \theta),
\end{equation}
where $A$, $\omega$, and $\theta$, denote the amplitude, frequency, and phase, respectively.

 To date, a ring lattice with a periodically modulated angular position has not been experimentally realized. Nevertheless, its key ingredients have been achieved independently. In particular, rotation sensing using shaken two-dimensional optical lattices has recently been demonstrated \cite{reinforcement_shaken_2d_giroscopio}, and ring lattice configurations with optical potentials rotating at constant angular velocity have also been realized experimentally  \cite{lattice_anillo_rota_lineal}.

 After an appropriate gauge transformation \cite{viebahn2020introduction}, the system's Hamiltonian takes the form,
\begin{equation}
    \widehat{H}_0 = \frac{\left[\hat{p}_\varphi - \mathcal{
A}(t)\right]^2}{2I_{rot}} + V(\hat{\varphi})\,,
\label{Hamiltoniano final single particle}
\end{equation}
where the potential satisfies $V\left(\varphi+\frac{2\pi}{N_s}\right) = V\left(\varphi\right)$, and $N_s$ denotes the number of lattice sites. The physical meaning of the gauge vector potential, $\mathcal{
A}(t) = I_{rot}[\Omega(t) - \dot\varphi_0(t)]$, is discussed below.

 In the presence of atomic interactions and for a lattice potential depth $V_0 \gtrsim 5E_R$, where $E_R$ is the recoil energy, the system is well-described by the tight-binding approximation. In this regime, the Hamiltonian reduces to the Bose-Hubbard model \cite{persistentcurrents-review},
\begin{equation}
    \hat H_{BH} = -J \sum_{l = 1}^{N_s} \left(e^{i\frac{\phi(t)}{N_s}}\hat a_l^{\dagger} \hat a_{l+1} + \text{h.c.}\right) + \frac{U}{2}\sum_{l=1}^{N_s} \hat{n}_l\left(\hat n_l -1\right)\,,
    \label{hamiltoniano bose hubbard completo}
\end{equation}
 where $J$ denotes the hopping energy and periodic boundary conditions are assumed for a lattice with  $N_s$  sites. The operator $\hat{a}_{l}^{\dagger}$ ($\hat{a}_l$) creates (annihilates) a particle at site $l$, and $\hat{n}_l=\hat{a}_{l}^{\dagger}\hat{a}_l$ is the corresponding particle number operator. The total number of particles, $N$, is then determined by the expectation value of the operator $\sum_{l=1}^{N_s} \hat{n}_l$.
The phase $\phi(t) = 2\pi \mathcal{A}(t)/\hbar$, known as the Peierls phase \cite{peierls1933theorie}, represents a fundamental tool for studying superfluidity in Bose-Einstein condensates (BECs) confined to optical lattices \cite{roth2003superfluidity,jaksch2003creation}. Consequently, several methods to engineer the Peierls phase in ultracold systems (including those driven by ac fields) have been proposed as a way to create artificial gauge fields \cite{Peierls-AC-Struck, Synthetic-magnetic-Nature-Spielman, Artificial-gauge-potentials-RMP-colloquium}. The last term in Eq.~(\ref{hamiltoniano bose hubbard completo}) describes the on-site atomic interaction whose strength $U$, which can be experimentally tuned via Feshbach resonances \cite{RMP-feshbach}.

 For a system subject to a constant angular acceleration $\alpha$, such that $\Omega(t) = \alpha t$, Eq.~(\ref{Eq:phase-shaking}) yields the following expression for the Peierls phase,
\begin{equation}
\frac{\phi(t)}{N_s} = \omega_B t + \tilde A\sin (\omega t + \theta),
\label{fase de peierls final}
\end{equation}
where $\tilde A \equiv A\left(\textstyle{2\pi I_{rot}\omega}\right)/\left(\textstyle{N_s\hbar}\right)$ is  the dimensionless driving amplitude, and 
\begin{equation}
    \omega_B = \frac{2\pi I_{rot}}{\hbar N_s}\alpha\, ,
    \label{Angular bloch frequency}
\end{equation}
is the angular Bloch frequency, whose precise measurement enables the estimation of the angular acceleration \cite{angularBOs}. Alternatively, the angular acceleration can be determined with higher precision by measuring average atomic supercurrents, as explained in the following section, where we discuss the working principle of the proposed atomtronic angular accelerometer.

\section{Supercurrents for inertial sensing}\label{sec:method}

 To understand the operation of an angular accelerometer based on the measurement of supercurrents in an angularly shaken ring lattice,  it is essential to characterize the atomic current induced by a time‑dependent Peierls phase. To this end, we quantify the atomic current by introducing the current operator, which characterizes the flow of atoms between neighboring sites of the ring lattice. It is defined as

\begin{equation}\label{Eq:current-operator}
\hat{\mathcal{J}} =\frac{1}{\hbar}\frac{\partial \hat{H}_{BH}}{\partial\phi}=-i\frac{\omega_J}{N_s} \sum_{l =1}^{N_s}\left(e^{i\frac{\phi(t)}{N_s}}\hat a^{\dagger}_l \hat a_{l+1} - \text{h.c.}\right),
\end{equation}
where $\omega_J=J/\hbar$ is the hopping frequency. The observable represented by $\hat{\mathcal{J}}$ can be measured experimentally by probing the phase
gradient between atoms trapped at neighboring sites \cite{cold-atoms-ring, stabilizing-persistent-currents-Nature}.

In this section, we analyze the response of this current in both the non‑interacting and weakly interacting regimes. We begin by deriving an analytical expression for the atomic current in the single‑particle limit, which provides physical insight into the sensing mechanism and serves as a benchmark for the many‑body simulations discussed later. We then investigate,  in both regimes, the non‑equilibrium dynamics induced by a time‑dependent Peierls phase in the Bose–Hubbard Hamiltonian [Eq.~\eqref{hamiltoniano bose hubbard completo}]. This protocol allows us to start from a well‑defined initial state with no imprinted phase. Specifically,  the system is initially prepared in the many‑body ground state, corresponding to a vanishing   $\theta=0$, in the Peierls phase.  Then, at $t=0$, the phase is quenched and subsequently evolves according to
\begin{equation}\label{eq.quench}
    \frac{\phi(t)}{N_s} =\begin{cases}
        0 &\text{if  } t = 0,\\
        \omega_Bt+\tilde A\sin(\omega t+\theta)&\text{if  } t > 0. \\
    \end{cases} 
\end{equation}

We first analyze the non‑interacting regime, where the current response admits a simple analytical description, as discussed in the following Subsection.

\subsection{Non-interacting regime}

In the single-particle limit ($U=0$), the expectation value of the current is found to be
 
\begin{equation}
    I(t) \equiv \langle \hat{\mathcal{J}}\rangle = I_0 \sin\left[\frac{\phi(t)}{N_s}\right]\,,
    \label{corriente igual sin de phi}
\end{equation}
where \(I_0 = \frac{2\;\omega_J}{ N_s}\) is the current amplitude. Here, we assume that the particle occupies the system’s ground state with zero momentum.

Combining Eqs.~(\ref{fase de peierls final}) and (\ref{corriente igual sin de phi}) and applying the Jacobi-Anger expansion, we obtain
\begin{equation}
\begin{aligned}
    \frac{I(t)}{I_0} =\sum_{n=-\infty}^{\infty}(-1)^nJ_{n}(\tilde A)\sin\left(\Delta\omega_nt - n\theta\right)\,,
\end{aligned}
\label{sumatoria final corriente}
\end{equation}
where $J_n(A)$ is the $n$-th Bessel function of the first kind and $\Delta\omega_n = \omega_B-n\omega$. Since the current contains multiple oscillatory contributions, an experimentally relevant sensing signal is obtained from the current averaged over a finite measurement time $\tau$,
\begin{equation}
\begin{aligned}
    \frac{\langle I\rangle_\tau}{I_0} =
    \sum_{n = -\infty}^{\infty}(-1)^nJ_n(\tilde A)\frac{1}{\tau}\int_0^\tau\sin(\Delta\omega_nt - n\theta)\,dt\,,
    \label{corriente promedio pre aprox}
\end{aligned}
\end{equation}
where $\langle \cdots\rangle_\tau$ denotes the time average over the interval $\tau$. When $\tau$ is sufficiently large, only the resonant term with $\Delta\omega_n = 0$ contributes significantly to the summation in Eq.~\eqref{corriente promedio pre aprox}, leading to peaks in the time-averaged current whenever the drive frequency satisfies, 
\begin{equation}
     \omega=\frac{\omega_B}{n}\equiv\omega_n.
\end{equation}

Therefore, when the drive frequency exactly matches the $m$-th subharmonic, the time-averaged current is approximately,
\begin{equation}
    \langle I\rangle_{\tau,m} \approx I_0 (-1)^{m+1}J_m(\tilde A)\sin(m\theta)\,.
    \label{amplitud corriente en funcion de A}
\end{equation}
This expression reveals that by tuning the drive amplitude $\tilde A$, one can selectively enhance or suppress the time-averaged current at specific subharmonics. Both the sign and magnitude of the current are governed by the Bessel function $J_m(\tilde A)$, enabling precise control over its direction and strength. Consequently, by sweeping the drive frequency until $\langle I\rangle_\tau$ becomes appreciable, one can directly determine the angular Bloch frequency within the resonance width, and hence the angular acceleration experienced by the atoms.

To gain further insight into the accuracy of measuring the Bloch frequency, we analyze the behavior of the time-averaged current for drive frequencies close to resonance with the angular Bloch frequency. Specifically, we consider frequencies near a subharmonic, i.e., $\omega \approx \omega_m$. In this regime, the harmonic component with small detuning $\Delta\omega_m$ varies slowly in time, whereas the remaining harmonics, with detunings $\Delta\omega_n$ for  $n \neq m$ oscillate rapidly. As a result, the dominant contribution to the time-averaged current arises from the $n=m$ term in Eq.~(\ref{corriente promedio pre aprox}). Thus, near the resonant subharmonic $\omega_m$, the time-averaged current is approximately,

\begin{equation}
\begin{aligned}
    \frac{\langle I\rangle_{\tau}\left(\Delta\omega_m\right)}{I_0} 
    \approx   (-1)^{m}J_m(\tilde A)\;{\rm sinc}\left(\frac{\Delta\omega_m \tau}{2}\right)\sin\left(\frac{\Delta\omega_m \tau}{2}-m\theta\right)\,,
    \label{corriente promedio pre sinc}
\end{aligned}
\end{equation}
which depends explicitly on the detuning $\Delta\omega_m $.

Equation~\eqref{corriente promedio pre sinc} reveals that the qualitative behavior of the time‑averaged current depends on the drive phase $\theta$. Of particular relevance for sensing is the case in which the phase satisfies
 $m\theta = (2k+1)\frac{\pi}{2}$, with $k \in \mathbb{Z}$. In this configuration, the averaged current reduces to a  sinc‑like resonance profile
\begin{equation}\label{current-sinc}
 \frac{\langle I\rangle_{\tau}\left(\Delta\omega_m\right)}{I_0} 
    \approx  (-1)^{m+k}J_m(\tilde A)\;{\rm sinc}\left(\Delta\omega_m \tau\right) ,
\end{equation}
which exhibits a maximum at the resonance condition and is therefore particularly well suited for precise frequency (and hence acceleration) estimation.

\begin{figure}
    \centering
    \includegraphics[scale=0.62]{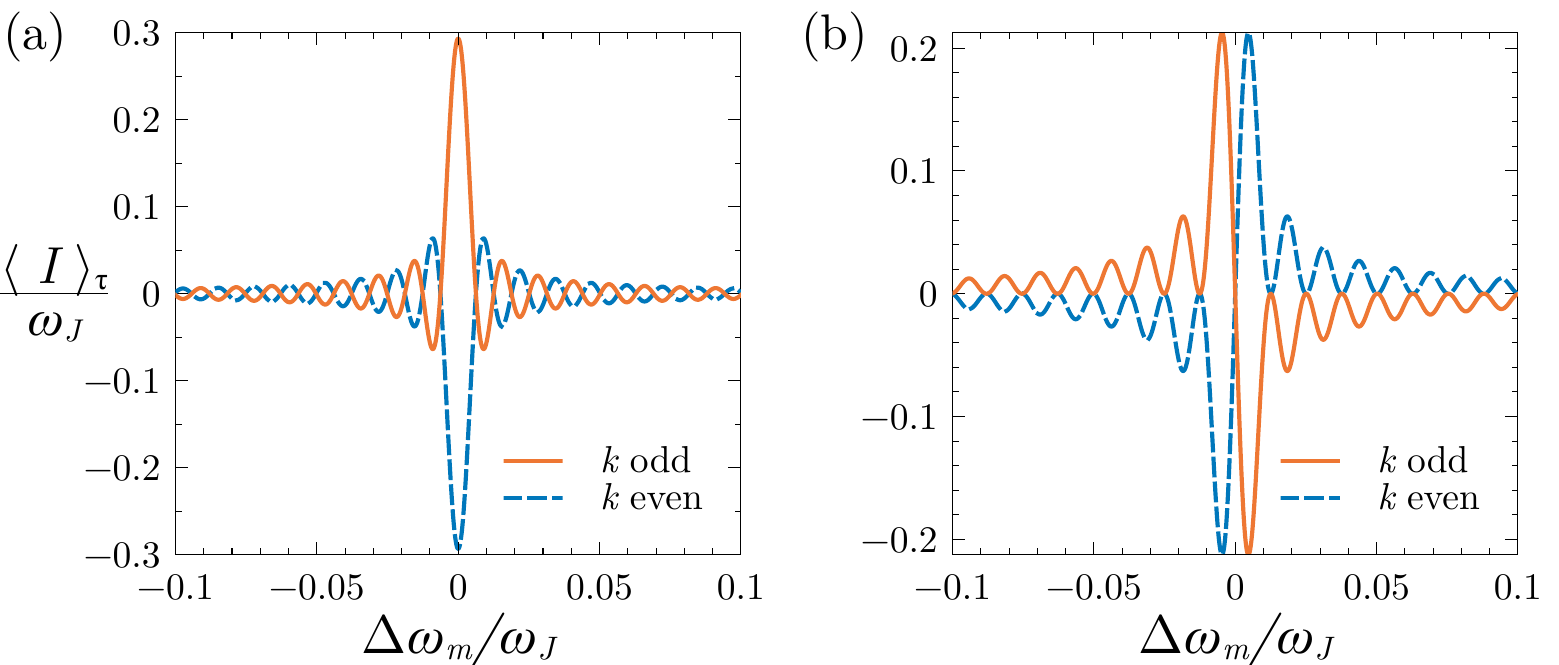}
    \caption{Time-averaged current as a function of $\Delta\omega_m/\omega_J$ near the resonance condition for (a) $m\theta =(2k+1)\frac{\pi}{2}$ and (b) $m\theta =k\pi$ . The remaining parameters are $\tau = 500/\omega_J$, $N=1$, $N_s = 3$, and $\tilde A=1$. 
    }
    \label{fig:corriente teorica cos}
\end{figure}

The time-averaged current as a function  $\Delta\omega_m$ is shown in Fig. \ref{fig:corriente teorica cos}(a).
As expected from Eq.~(\ref{current-sinc}), the time-averaged current is modulated by the sinc function, with the width of the peak around resonance inversely proportional to the measurement time $\tau$. Consequently, an efficient measurement strategy is to start with a relatively short $\tau$, producing a broader resonance peak that facilitates detection during a frequency sweep. Increasing $\tau$ subsequently narrows the peak, enhancing the precision in determining the angular Bloch frequency and, in turn, the angular acceleration. However, the maximum reachable $\tau$ is limited by the condensate coherence time, which in turn constrains the precision achievable in the non-interacting regime.

For completeness, we note that when the phase satisfies $m \theta = k\pi$, with $k \in \mathbb{Z}$, the time‑averaged current takes the form:
\begin{equation}\label{current-sinc_2}
 \frac{\langle I\rangle_{\tau}\left(\Delta\omega_m\right)}{I_0} 
    \approx  (-1)^{m+k}J_m(\tilde A)\;\frac{1}{\tau\, \Delta\omega_m}{\rm sin^2}\left(\frac{\Delta\omega_m \tau}{2}\right)\ .
\end{equation}
This expression is antisymmetric about the resonance and vanishes exactly at $\Delta \omega_m=0$, as shown in Fig. \ref{fig:corriente teorica cos}(b).
Since this configuration does not yield a detectable supercurrent at resonance, it is not suitable for sensing applications.

Overall, the dependence of the time-averaged current in Eq.~(\ref{corriente promedio pre sinc}) on the static phase $\theta$, the drive amplitude, and the resonant subharmonic index allows for fine control over the current at different resonant frequencies. This enables selective enhancement of the peak associated with a given subharmonic, while providing detailed insight into the net atomic current in the vicinity of resonance under well-controlled conditions.

\begin{figure}
    \centering
    \includegraphics[scale=0.6]{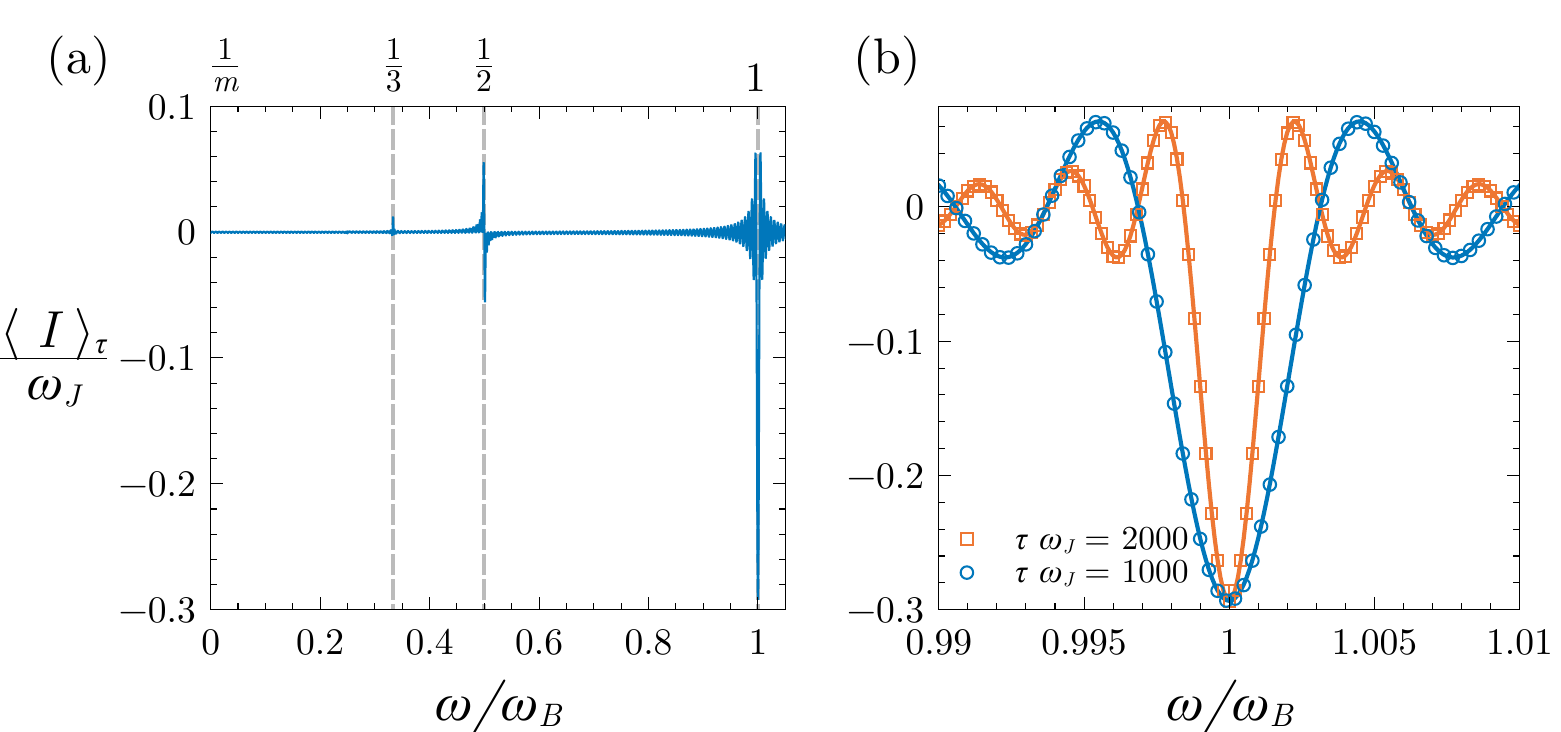}
    \caption{(a) Time-averaged current as a function of $\omega/\omega_B$, for $\tau = 1000/\omega_J$. 
(b) Time-averaged current near the resonance at $\omega/\omega_B = 1$ for measurement times $\tau = 1000/\omega_J$ and $\tau = 2000/\omega_J$. 
The numerical simulation results are shown as blue empty circles for $\tau = 1000/\omega_J$ and orange empty squares for $\tau = 2000/\omega_J$, while the corresponding theoretical predictions from Eq.~(\ref{current-sinc}) are shown as solid lines with matching colors. 
The analytical predictions are in excellent agreement with the numerical simulations for $U/J = 0$. 
The remaining parameters are $N = 1$, $N_s = 3$, $\tilde A = 1$, and $\theta = \pi/2$.}
    \label{fig:resonance-peaks}
\end{figure}

\begin{figure}
    \centering
    \includegraphics[scale=0.28]{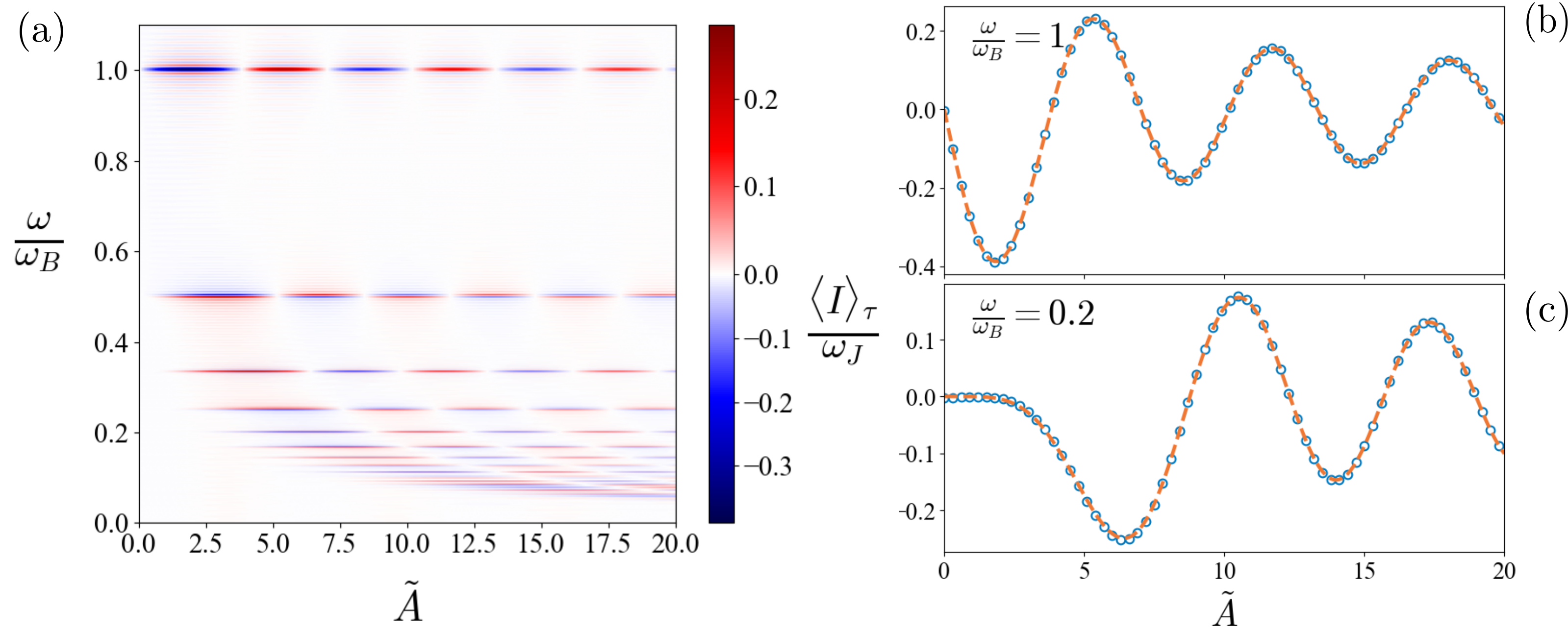}
    \caption{(a) Numerical time-averaged current as a function of $\omega/\omega_B$ and $\tilde{A}$ for $N=1$, $N_s=3$, $\theta = \pi/2$, $U = 0$, and $\tau = 500/\omega_J$. (b),(c) Numerically computed time-averaged current at resonance as a function of $\tilde{A}$ for $\omega/\omega_B = 1$ and $\omega/\omega_B = 0.2$ (empty dots), together with the corresponding theoretical prediction from Eq.~\eqref{amplitud corriente en funcion de A} (dashed orange line), showing excellent agreement.}
    \label{fig:heatmap corriente vs w vs A y curvas}
\end{figure}

\subsubsection*{Non-equilibrium dynamics induced by a phase quench}
While the above analysis assumes a system with an initially imprinted phase, realistic experimental preparations start from a non-transporting ground state. Consequently, when the phase quench defined in Eq.~(\ref{eq.quench}) is applied, the system is driven into non‑equilibrium dynamics. To capture the essential features of the current response induced by such a phase quench, it is sufficient to consider a three-site ring lattice \cite{morales2012-trimer,triangular-trimer-bose-hubbard}, which constitutes the minimal system with periodic boundary conditions. The resulting trimer therefore provides the most elementary model on which our atomtronic angular accelerometer could be tested.  

The numerical results for the time-averaged current as a function of the driving frequency for $\tilde A = 1.0$ and $\theta = \pi/2$, are shown in Fig.~\ref{fig:resonance-peaks}. The time-averaged current diminishes and becomes negligible as the frequency moves away from integer divisors of $\omega_B$, as illustrated in Fig.~\ref{fig:resonance-peaks}(a). Near the resonance frequencies, the behavior associated with the subharmonics $m=1,3$ and $m=2$ closely follows the predictions of Eqs.~(\ref{current-sinc}) and (\ref{current-sinc_2}), respectively. A close-up of the time-averaged current near the main resonance ($\omega/\omega_B = 1$) is shown in Fig.~\ref{fig:resonance-peaks}(b) for two measurement times  $\tau =1000/\omega_J$ and $\tau=2000/\omega_J$, where numerical simulations and analytical predictions show excellent agreement. This agreement persists despite the fact that the numerical simulations include a phase quench. Such a result is not surprising, since in the single-particle limit the resonance current response is governed solely by frequency matching and is therefore insensitive to whether the Peierls phase is introduced adiabatically or through a sudden quench. As expected from Eq.~(\ref{current-sinc}), the results display a sinc-like profile, featuring a pronounced net atomic current near the resonance. Moreover, Fig.~\ref{fig:resonance-peaks}(b) clearly illustrates that the resonance width—and, consequently, the sensitivity of the time-averaged current measurement—depends on the measurement time $\tau$.

To provide a more comprehensive view of the resonance behavior across the parameter space, Fig.~\ref{fig:heatmap corriente vs w vs A y curvas}(a) presents the time-averaged current as a function of both the driving frequency and amplitude. The figure shows that tuning $\tilde{A}$ enables both suppression and enhancement of the time-averaged current at different subharmonics, and even allows control over the direction of the net atomic current. When combined with the choice of $\theta$, this enables precise control over which subharmonics are suppressed or activated, thereby facilitating their selective detection as the driving frequency is varied. Furthermore, Figs.~\ref{fig:heatmap corriente vs w vs A y curvas}(b)-(c) clearly highlight the resonance peak amplitudes at $\omega/\omega_B = 1$ and $\omega/\omega_B = 0.2$, demonstrating an excellent agreement between the numerical simulations (empty dots) and the predictions of Eq.~\eqref{amplitud corriente en funcion de A} (dashed lines).

Despite the high degree of control and tunability, the dependence of the resonance width on the measurement time remains a fundamental limitation on the precision of angular acceleration measurements. This Fourier‑limited sensitivity is well known in Bloch‑oscillation‑based linear accelerometers \cite{blochoscillationsgravity, tarallo2012resonanttunnelingBO}, and it also arises in theoretical proposals for ultracold-atom-based angular accelerometers \cite{angularBOs}. In the following Subsection, we demonstrate that this limitation can be overcome by incorporating weak interatomic interactions into the system.

\subsection{Weakly interacting regime}
Thus far, our analysis has focused on the single‑particle limit, which captures the basic sensing mechanism and allows for a transparent analytical treatment. In realistic Bose–Einstein condensates, however, interatomic interactions are inevitably present and can significantly influence the precision of time‑averaged current measurements. Understanding how weak interactions modify the resonance structure of the current is therefore essential for assessing the performance of current‑based inertial sensing schemes.

\begin{figure}
    \centering
    \includegraphics[scale=0.54]{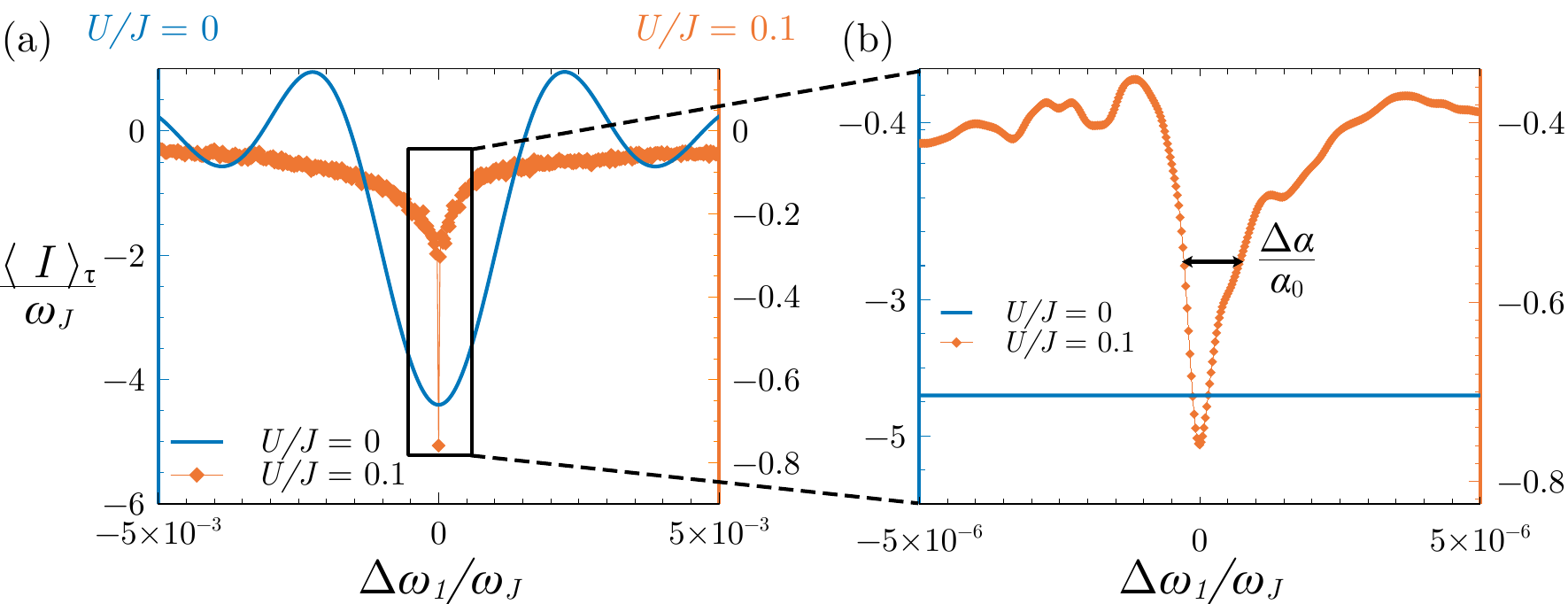}
    \caption{(a) Time-averaged current as a function of $\Delta\omega_1/\omega_J$ for $\theta=\pi/2$ and $\tau = 2000/\omega_J$. The non-interacting case (solid blue line) is compared with the weakly interacting case (orange diamonds) for $N_s=5$, $N=15$, and $U/J=0.1$. The left (right) $y$-axis corresponds to the non-interacting (interacting) data, with colors matching their respective curves. The resonance width is significantly narrower in the interacting case. (b) Zoom-in of the resonance region. The peak corresponding to $U/J= 0.1$ has an appreciable width, while the peak for $U=0$ appears nearly flat on this scale. The double-headed arrow marks the full width at half maximum (FWHM), $\Delta \alpha/\alpha_0$, which we use as a measure of the device sensitivity. As in panel (a), the left (right) $y$-axis corresponds to the non-interacting (interacting) data.}
    \label{fig:corriente N1 vs N13}
\end{figure}
To systematically explore the effects of interatomic interaction, we study the nonequilibrium dynamics following the phase quench Eq.~(\ref{eq.quench}) by performing numerical simulations of a ring lattice with multiple particles in the weakly interacting regime  $U/J \ll 1$. Experimentally, this regime is readily accessible, as the interaction strength $U$ can be tuned via Feshbach resonances \cite{RMP-feshbach}, while the tunneling amplitude $J$ can be controlled by adjusting the lattice depth. The numerical simulations were carried out for systems with $N_s = 3,\,4 \text{ and }5$ sites, and filling factors $\nu=N/N_s=1,2, \text{ and }3$, enabling a systematic analysis of weak interaction effects and allowing us to benchmark the results against the non-interacting predictions discussed in the previous subsection.

Fig.~\ref{fig:corriente N1 vs N13}(a) shows the time-averaged currents as a function of the normalized driving frequency, $\omega/\omega_B$, for five-site rings with $N = 15$ for $U/J = 0$ (solid blue line) and $U/J=0.1$ (orange diamonds), both evaluated at $\tau = 2000/\omega_J$. Complementary, a zoom-in of the resonant peak is shown in Fig.~\ref{fig:corriente N1 vs N13}(b). 

According to Eq.~(\ref{Angular bloch frequency}), the sensitivity for measuring the angular acceleration is given by
$\Delta\alpha = \Delta\omega_B (\hbar N_s)/(2\pi I_{\mathrm{rot}})$ and is therefore ultimately limited by the uncertainty in the angular Bloch frequency, $\Delta\omega_B$. The relative uncertainty $\Delta\omega_B/\omega_J$ can be extracted from the full width at half maximum (FWHM) of the time-averaged supercurrent resonance, as illustrated in Fig.~\ref{fig:corriente N1 vs N13}(b). One thus obtains,
\begin{equation}
\mathrm{FWHM} = \frac{\Delta\omega_B}{\omega_J} = \frac{\Delta\alpha}{\alpha_0},
\end{equation}
where $\alpha_0 = N_s \hbar \omega_J/(2\pi I_{\mathrm{rot}})$. Consequently, the FWHM of the time-averaged supercurrent resonance provides a direct measure of the accelerometer sensitivity ($\Delta\alpha=\alpha_0\;{\rm FWHM}$), with narrower resonance peaks allowing for a more precise determination of the angular Bloch frequency and, hence, of the angular acceleration. Note, however, that smaller values of $\Delta\alpha/\alpha_0$ correspond to higher sensitivity, as they reflect improved measurement precision.

The results shown in Fig.~\ref{fig:corriente N1 vs N13} indicate that, for weakly interacting atoms, the resonance peak becomes significantly narrower than in the non-interacting case, thereby enhancing the precision of angular acceleration measurements. However, the presence of weak interactions can also reduce the peak amplitude of the time-averaged current, as shown in Fig.~\ref{fig:corriente N1 vs N13}. Thus, while interactions enhance sensitivity, the reduced current may pose experimental challenges, requiring a trade-off that defines the interaction range for optimal operation, as discussed later in this article.

\subsubsection*{Resonance peak narrowing via dephasing-induced interference}
\begin{figure}
    \centering
    \includegraphics[scale=0.4]{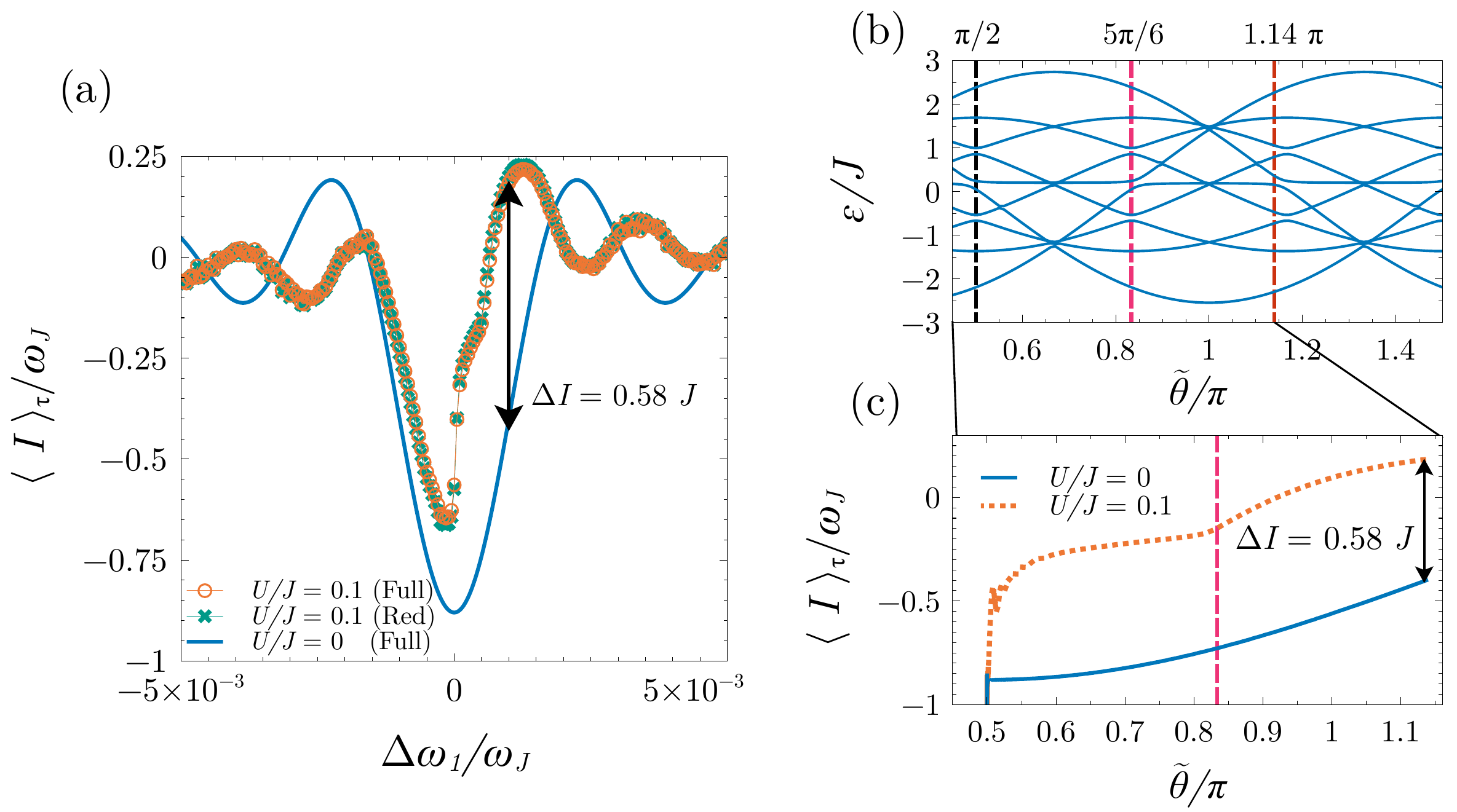}
    \caption{(a) Time-averaged current as a function of the detuning $\Delta \omega_1/\omega_J$ for $\theta=\pi/2$ and $\tau = 2000/\omega_J$. The weakly interacting case is shown both from full simulations (orange empty circles) and from the reduced Hamiltonian (green crosses) for $N=3$ and $N_s=3$, together with the non-interacting result (solid blue line). The vertical double-headed arrow indicates the difference $\Delta I = 0.58\,J$ between the interacting and non-interacting time-averaged currents, evaluated at $\Delta \omega_1/\omega_B = 10^{-3}$. (b) Quasienergy spectrum of the reduced Hamiltonian [Eq.~(\ref{reduced-Hamiltonian})] for $U/J=0.1$. The magenta dashed vertical line marks the avoided crossing at $\tilde{\theta}=5\pi/6$. 
    (c) Time-averaged current as a function of $\tilde{\theta}$ for $U/J=0$ (solid blue line) and $U/J=0.1$ (orange dotted line), computed at fixed detuning $\Delta \omega_1/\omega_B = 10^{-3}$. The magenta dashed vertical line indicates the same value $\tilde{\theta}=5\pi/6$, corresponding to the avoided crossing shown in panel (b). The same difference $\Delta I$ is highlighted by a vertical double-headed arrow.}
    \label{fig:effective}
\end{figure}

The narrowing of the resonance peak observed in the net atomic current can be understood by analyzing the effective dynamics derived from the ac-driven Bose-Hubbard Hamiltonian (see Appendix \ref{apenA} for details). For illustration, we consider the case $\tilde{A} = 1$ and small detuning, $\Delta \omega_1 \ll \omega_B$. Under these conditions, the time evolution is dominated by the $n=0,1$ modes, and the full Hamiltonian can effectively be reduced to  
\begin{equation}\label{reduced-Hamiltonian}
 H_{red}= -J J_0(\tilde{A})\sum_{l = 1}^{N_s}\left(e^{i\omega_B t }\hat{a}^{\dagger}_{l+1}\hat{a}_{l}+\text{h.c.}\right)+JJ_1(\tilde{A}) \sum_{l = 1}^{N_s}\left(e^{i\tilde{\theta}}\hat{a}^{\dagger}_{l+1}\hat{a}_{l}+\text{h.c.}\right) + \frac{U}{2}\sum_{l=1}^{N_s} \hat{n}_l(\hat{n}_l-1).
\end{equation}
where $\tilde{\theta}=\Delta \omega_1 t - \theta$.
To verify the validity of this reduced Hamiltonian,  we consider a ring with $N_S=N=3$. Fig.~\ref{fig:effective}(a) shows the averaged current near resonance for both the full system and the reduced model, revealing excellent agreement between the two descriptions.
 In this near-resonant regime, the system can be averaged over the period $T_B=2\pi/\omega_B$, namely the micromotion time scale \cite{weitenberg2021tailoring,eckardt2017colloquium,goldman2014periodically}, allowing us to compute the quasienergies as functions of the parameter   $\tilde{\theta}$ shown in Fig.~\ref{fig:effective}(b). The resulting quasienergy spectrum, shown in Fig.~\ref{fig:effective}(b), exhibits multiple avoided crossings, with gap sizes proportional to the interaction strength $U$.

 The presence of a finite detuning introduces a “rate” of change of $\tilde{\theta}$, corresponding to a displacement along the quasienergy bands, so that adiabatic following is repeatedly interrupted at the avoided crossings. This leads to interference among different Floquet modes, each associated with a distinct mean momentum. Consequently, destructive interference accumulates over time, suppressing the net current. This interpretation is supported by the cumulant of the average current shown in Fig.~\ref{fig:effective}(c), where pronounced changes in the current response occur at the avoided crossings at $\tilde{\theta} = \pi/2$ (vertical black dashed line) and $\tilde{\theta} = 5\pi/6$ (vertical magenta dashed line).

 Similar interference mechanisms, previously studied in tilted lattices with interacting atoms \cite{dephasing,witthaut2005bloch,kolovsky2004bloch}, have been shown to cause dispersion, dephasing, and delocalization due to interaction-induced coupling between Floquet modes. Although those studies did not explicitly analyze the current, our results demonstrate that the same interference processes directly account for the observed narrowing of the resonance peak. Moreover, the pattern of fluctuating current values near resonance is consistent with this interpretation, as shown in Fig.~\ref{fig:corriente N1 vs N13}. 

\subsubsection*{Sensitivity enhancement via interactions}

To quantify the impact of weak interactions on the precision of angular acceleration measurements, we analyze the resonance peak width as a measure of the sensitivity. Specifically, we extract the full width at half maximum (FWHM) of the time-averaged current $\Delta \alpha/\alpha_0$, as illustrated in Fig. \ref{fig:corriente N1 vs N13}. This allows us to systematically compare how the resonance sharpens with increasing interaction strength. 

\begin{figure}
    \centering
    \includegraphics[scale=0.49]{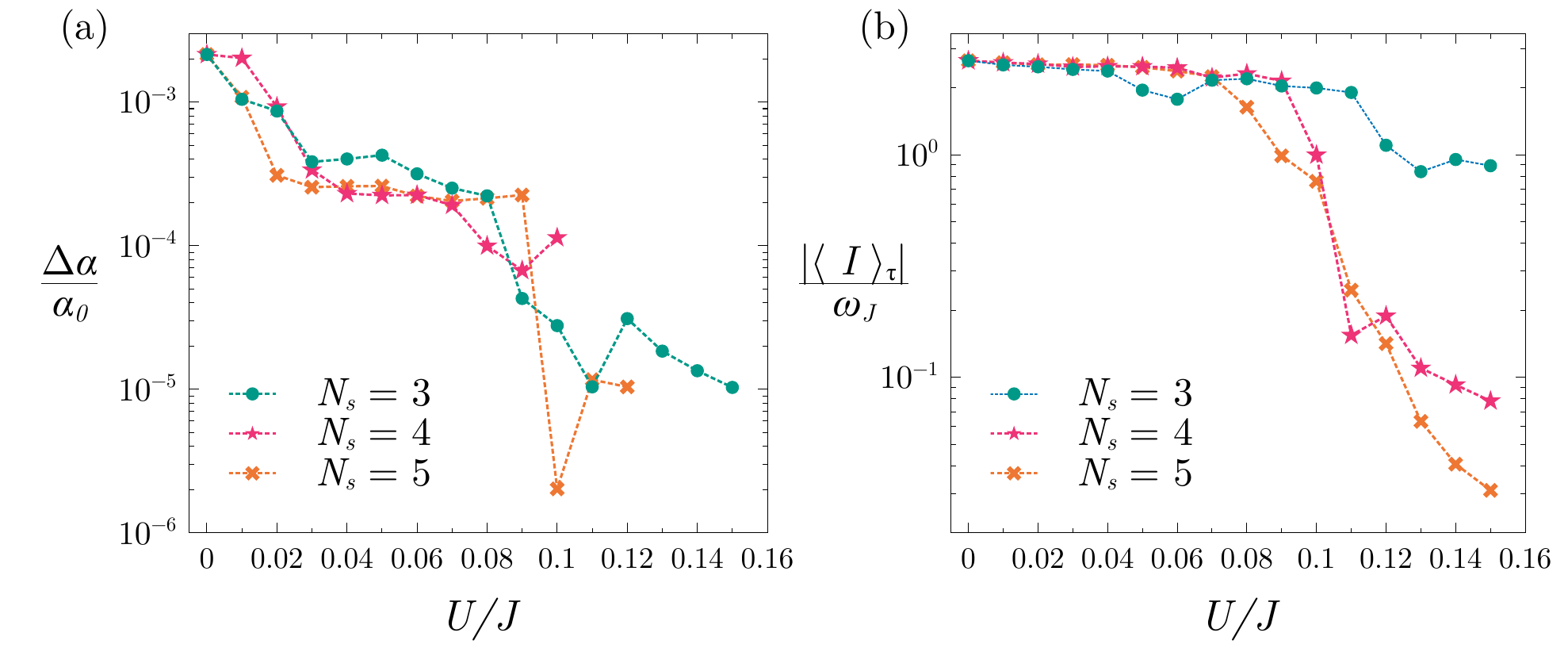}
    \caption{(a) Estimated angular-acceleration sensitivity (in units of $\alpha_0$) as a function of interaction strength $U/J$.
(b) Absolute value of the time-averaged current (in units of $\omega_J$) at the main resonant peak, shown as a function of $U/J$. In both panels, results are presented for three system sizes: $N_s = 3$ (green circles), $N_s = 4$ (magenta stars), and $N_s = 5$ (orange crosses). Simulations are performed at the main resonance, $\omega/\omega_B = 1$, with $\tau = 2000/\omega_J$, $\theta = \pi/2$, $\tilde{A} = 1$, and  a filling factor $\nu = 3$.}
    \label{fig:delta alpha vs U}
\end{figure}

Fig.~\ref{fig:delta alpha vs U}(a) shows the sensitivity as a function of the interaction parameter $U/J$ in the weakly interacting regime ($U/J\ll 1$). We consider various system sizes, all with a filling factor of three, which enables high sensitivity values. In all cases, angular acceleration sensitivity is substantially enhanced, with the resonance peak narrowing by up to two orders of magnitude compared to the non-interacting (single-particle) scenario. This narrowing corresponds to a dramatic theoretical improvement in the sensitivity of the proposed atomtronic angular accelerometer.

However, increasing the interaction strength also leads to a suppression of the current.  Fig.~\ref{fig:delta alpha vs U}(b) shows the current at the resonance point as a function of $U/J$, revealing that the current begins to decay around $U/J=0.1$, and eventually drops by two orders of magnitude for larger $U/J$ values.
To qualitatively understand this behavior, we derive in Appendix~\ref{apenB} an approximate expression for the time‑averaged current near the $m$-th resonance,

\begin{eqnarray}\label{eqn:current-k-representation}
    & &\frac{\langle I\rangle_{\tau}^{U}}{I_0}=\frac{\langle I\rangle_{\tau}}{I_0}\left(N-\frac{1}{2}\sum_{k\neq 0}\frac{\omega_U^2}{\gamma_{km}^2}\right)\\
    &+&\frac{(-1)^m}{4}J_m(\tilde{A})\sum_{k\neq 0}\sum_{s=\pm1}\left\{\frac{\omega_U^2}{\gamma_{km}^2} {\rm sinc}\left[\frac{(\Delta\omega_m+2s\gamma_{km})\tau}{2}\right]\sin\left[\frac{(\Delta\omega_m+2s\gamma_{km})\tau}{2}-m\theta\right]\right\},\nonumber
\end{eqnarray}
where  $\gamma_{km}=\sqrt{(\Omega_{km}+\omega_U)^2-\omega_U^2}$ 
with $\Omega_{km}=2\omega_J\left[1-(-1)^m J_m(\tilde{A})\cos(k)\cos(m\theta)\right]$ and $\omega_U=U N/(\hbar N_s)$.

To gain insight into the predictions of Eq.~(\ref{eqn:current-k-representation}), we first examine the resonant case $\Delta\omega_m = 0$. In the non-interacting limit, only the leading contribution survives, yielding a current that scales linearly with $N$. In the presence of interactions, however, additional terms with $k \neq 0$ appear and give rise to a negative correction proportional to $\omega_U^2$, thereby reducing the total current. Physically, this suppression originates from interaction-induced quasiparticle excitations, which deplete the condensate population. This provides a qualitative explanation for the monotonic decrease in the current observed in Fig.~\ref{fig:delta alpha vs U}(b).

Away from resonance ($\Delta\omega_m\neq 0$), the interaction‑induced sideband terms become oscillatory components with frequencies shifted by $\pm 2\gamma_{km}$. These sidebands are generally out of phase with the primary resonant term, resulting in partial cancellation of the current. This additional destructive interference further suppresses the net signal. We emphasize that the expression above includes only the dominant sideband contributions; higher‑order terms, not captured in this expression, would introduce further oscillatory components that can reinforce the overall interference pattern.

From an experimental perspective, the suppression of the current poses a challenge for signal detection, placing a practical limit on the range of interaction strengths suitable for high-sensitivity measurements of angular acceleration. This indicates the existence of an optimal operational regime, where a measurable current is maintained while preserving high sensitivity. Outside this regime, either the current is too small to detect, or the sensitivity approaches that of the non-interacting case.

\begin{figure}
    \centering
    \includegraphics[scale=0.33]{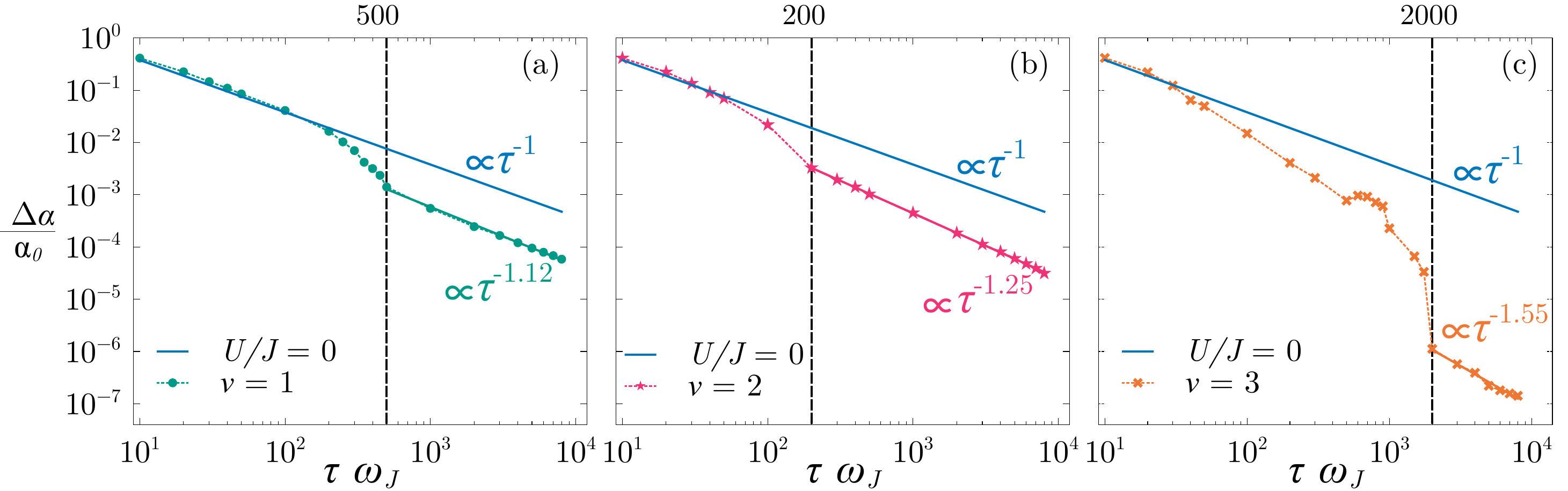}
\caption{Estimated angular-acceleration sensitivity as a function of the measurement time (in units of $1/\omega_J$) on a log-log scale. Panels (a), (b), and (c) correspond to integer filling factors $\nu = 1$, $\nu = 2$, and $\nu = 3$, respectively, and share the same vertical axis. In each panel, the sensitivity obtained within the analytical non-interacting approximation is shown as a solid blue line, while numerical results for $N_s = 5$ and $U/J = 0.1$, are shown as green dots in (a), magenta stars in (b), and orange crosses in (c). Power-law fits in the long-time regime, where a linear trend emerges in the log-log representation, are shown as solid lines.}
    \label{fig:delta alpha vs tau}
\end{figure}

\subsection{Maximal sensitivity values and experimental realization}

In this section, we analyze the sensitivity of angular-acceleration measurements as a function of the averaging time, accounting for atomic interactions.

Figure~\ref{fig:delta alpha vs tau} compares the sensitivity in the weakly interacting regime for filling factors $\nu=1,2\;\ \text{and}\,\ 3$ with that of the single-particle case. In the non-interacting limit (blue solid line), the resonance width scales inversely with the averaging time, reflecting the Fourier limit. This constraint is common to ac-driven analog sensors, including Bloch-oscillation-based accelerometers~\cite{blochoscillationsgravity,tarallo2012resonanttunnelingBO}.

When interactions are included, the sensitivity exhibits a pronounced enhancement of the scaling, due to interaction‑induced resonance narrowing.  Figure~\ref{fig:delta alpha vs tau} shows that, as the averaging time increases—allowing sufficient time for interatomic correlations to develop—the sensitivity progressively deviates from the Fourier-limited scaling $\propto \tau^{-1}$. This trend is illustrated in panels (a), (b), and (c), corresponding to filling factors $\nu=1,2,$ and $3$, respectively. The deviation from Fourier scaling occurs at shorter
averaging times for larger filling factors, indicating a stronger interaction-induced enhancement associated with increasing many-body contributions. This behavior is expected, since
the narrowing of the resonance width originates from the interference of sideband harmonics
in the current response [Eq.~(\ref{eqn:current-k-representation})], whose cumulative contribution becomes increasingly pronounced at longer averaging times.

A further distinctive feature across the three panels is that, beyond a filling‑dependent crossover time, the sensitivity approaches an asymptotic power‑law scaling of the form
\begin{equation}
    \Delta \alpha/\alpha_0 \propto \tau^{-p}
\end{equation}
with exponents $p\simeq 1.12,1.25,$ and $1.55$ for $\nu=1,2,3$, respectively, and $U/J=0.1$. This scaling identifies a regime dominated by interaction‑driven many‑body dynamics, where the resonant processes accumulated during the time evolution govern the ultimate 
sensitivity. Remarkably, by the end of the crossover time (marked by the vertical dashed lines in Fig.~\ref{fig:delta alpha vs tau}), the system has fully transitioned from the initial regime with approximate Fourier-limited scaling ($p=1$) to the interaction-driven regime characterized by $p > 1$. By this point, the sensitivity is already substantially enhanced (by roughly three orders of magnitude for $\nu = 3$). Beyond the crossover, the interaction-induced scaling ($p > 1$) further improves the sensitivity with increasing averaging time.

This regime contrasts sharply with the non-interacting case, which strictly follows the Fourier-limited $\tau^{-1}$  scaling (blue line). The strong dependence of the scaling exponent on the filling factor indicates that interactions give rise to distinct dynamical regimes, which can be exploited to achieve substantially higher sensitivities at shorter measurement times. This feature is particularly advantageous in realistic cold-atom experiments, where finite coherence times and technical noise impose stringent constraints on measurement duration and stability.

To assess the performance of our proposed accelerometer relative to previous designs, we compare our sensitivity results with the theoretical predictions of Ref.~\cite{angularBOs} for a non-driven lattice. In our calculations, we use $\omega_J \approx 500~\mathrm{Hz}$ (corresponding to a lattice depth $V_0 \approx 7E_R$~\cite{Rubio-Abadal-2020:Phys.Rev.X}, where $E_R$ is the recoil energy) and $N_s = 5$, whereas Ref.~\cite{angularBOs} considers $N_s = 20$. For consistency in the comparison, we adopt  a lattice radius of $90~\mu\mathrm{m}$ and atoms of mass corresponding to ${}^{87}\mathrm{Rb}$ as in Ref.~\cite{angularBOs}. For these parameters, Ref.~\cite{angularBOs} predicts a sensitivity of $\Delta\alpha \approx 10^{-4}~\mathrm{rad/s^2}$ for a measurement time $\tau = 10~\mathrm{s}$ using $N = 10^5$ atoms. By contrast, our fully quantum, weakly interacting model achieves $\Delta\alpha \approx 8 \times 10^{-6}~\mathrm{rad/s^2}$ with only $N = 15$ atoms for the same measurement time. This represents a significant improvement in sensitivity, while reducing the required particle number by four orders of magnitude, demonstrating a substantial enhancement in measurement precision. Furthermore, the lattice-shaking amplitude in our model is given by

\begin{equation}
A^* = \frac{N_s \hbar}{\pi I_{\rm rot}} \tilde{A},
\end{equation}
which can be experimentally tuned by adjusting the number of lattice sites $N_s$, providing a convenient control knob for optimizing the sensor’s performance.

Overall, the results discussed above underscore the novelty and potential of our proposed atomtronic angular accelerometer, demonstrating that weak interactions can be harnessed to surpass Fourier-limited sensitivity and achieve high-precision inertial sensing with only a few atoms.

\section{Concluding remarks}\label{sec:concl}

We have proposed and theoretically analyzed an atomtronic angular accelerometer based on an angularly ac-shaken ring lattice. The sensing signal is the time-averaged atomic current, which encodes the external angular acceleration. In the single-particle regime, we demonstrated—both analytically and numerically—that a nonzero average current emerges when the drive frequency matches a submultiple of the angular Bloch frequency; for other frequencies, the current remains negligibly small provided the averaging time is sufficiently long. Near resonance, the response exhibits a sinc-like profile with a Fourier-limited width, while the drive phase controls the direction of the atomic current.

Our analysis further shows that weak atomic interactions can fundamentally enhance sensitivity beyond the conventional Fourier limit by significantly narrowing the resonance. In ring-lattice systems, even weak interactions reduce the resonance width by several orders of magnitude compared to the non-interacting case. In addition, the interacting regime displays a markedly steeper dependence of the sensitivity on the averaging time than in the Fourier-limited scenario. This pronounced dependence on particle number reveals distinct interaction-driven dynamical regimes that can be harnessed for quantum-enhanced metrological performance.

We identified the physical origin of the resonance narrowing as dephasing-induced interference among Floquet modes. Using a reduced Hamiltonian valid near resonance, we showed that the quasienergy spectrum develops multiple avoided crossings, with gap sizes set by the interaction strength. A finite detuning drives the system along the quasienergy bands, causing repeated breakdowns of adiabatic following and generating interference between modes with different mean momenta. The resulting destructive interference accumulates over time, suppressing the net current and thereby narrowing the resonance. This mechanism is consistent with the observed sideband structure and with the interaction-dependent corrections captured by our momentum-space expression for the averaged current.

Finally, we showed that the metrological advantage introduced by interactions is accompanied by a reduction of the resonant current signal. As the interaction strength increases within the weakly interacting regime, the current decreases and can be suppressed by up to two orders of magnitude. Despite this trade-off, our proposal achieves superior sensitivity compared to mean-field predictions~\cite{angularBOs}. Moreover, the scheme is experimentally feasible with current ultracold-atom platforms, where interactions can be tuned via Feshbach resonances and the lattice-shaking amplitude can be controlled through the lattice laser parameters.

In summary, we have introduced an atomic‑current‑based inertial sensor whose sensitivity is enhanced by weak atomic interactions. We demonstrated Fourier‑beating scaling exponents and identified the interference mechanism responsible for the interaction‑induced resonance narrowing, thereby establishing a route toward few‑atom, high‑precision inertial sensing. These results pave the way for next‑generation atomtronic inertial sensors and may stimulate further theoretical and experimental exploration of interaction‑enabled quantum metrology in lattice‑based platforms.

\section{Acknowledgments}
 S. Carmona-López acknowledges financial support from ANID through the National Master’s Scholarship, ANID–Subdirección de Capital Humano/Magíster Nacional/2025–22260327. F. Isaule acknowledges funding from ANID through FONDECYT Postdoctorado 3230023. L.M-M acknowledges financial support from the Institute of Physics, PUC.

\appendix
\section{Reduced Hamiltonian}\label{apenA}

The Bose-Hubbard Hamiltonian (\ref{hamiltoniano bose hubbard completo}) can be recast using the Jacobi-Anger expansion as
\begin{equation}\label{suma-Hamiltonianos}
H= \sum_{n=-\infty}^{\infty} H_0^{(n)}+\frac{U}{2}\sum_{l=1}^{N_s} \hat{n}_l(\hat{n}_l-1),
\end{equation}
where
\begin{equation}
H_0^{(n)}= \Omega_n \sum_{l = 1}^{N_s} \left(e^{i(\Delta \omega_n t - \chi_n)}\hat{a}^{\dagger}_{l+1}\hat{a}_{l}+\text{h.c.}\right)
\end{equation}
with $\Omega_n=(-1)^{n+1}JJ_n(\tilde{A})$ and $\chi_n=n\theta$.
Hence, by tuning the amplitude of the driving $\tilde{A}$, the amplitude  $\Omega_n$ of the Hamiltonian terms $H_0^{(n)}$ can be controlled.

Moreover, Hamiltonians $H_0^{(n)}$ with large detunings $\Delta \omega_n$, oscillate rapidly and therefore can be neglected for a long term dynamics. Thus, for $\tilde{A}=1$ and small detuning $\Delta \omega_1\ll  \omega_B$,  the hopping term Hamiltonian can be reduced to two terms, i.e.  
\begin{equation}
\sum_{n=-\infty}^{\infty} H_0^{(n)}\approx  H_0^{(0)}+H_0^{(1)}\approx -J \sum_{l = 1}^{N_s}\left[ \left(J_0(\tilde{A})e^{i\omega_B t }-J_1(\tilde{A})e^{i(\Delta \omega_1 t - \theta)}\right)\hat{a}^{\dagger}_{l+1}\hat{a}_{l}+\text{h.c.}\right].
\end{equation}

Since $\Delta\omega_1 \ll \omega_B$, the evolution operator over one
Bloch period $T_B$ may be approximated by
\begin{equation}
U(T_B)
\simeq
{\cal T}\exp\!\left[
-i\!\int_0^{T_B} \left( H_0^{(0)}(t)+H_0^{(1)}(t)\right) dt
\right].
\end{equation}

\section{Bogoliubov Approximation}\label{apenB}

Assuming that all particles initially occupy the zero-momentum condensate state and adopting the mean-field approximation, the Bose–Hubbard Hamiltonian in momentum space can be written as
\begin{equation}
\hat{H}_{MF} = E_{MF} + \hat{H}_{B},
\end{equation}
where $E_{MF}$ denotes the condensate energy and
\begin{equation}\label{eqn:h-bogoliubov}
\hat{H}_{B} = \sum_{k} \xi_k(t)\bdk{k} \bk{k}
+ \frac{U N}{2N_s} \sum_{k} \left( \bdk{k} \bdk{-k} + \mathrm{h.c.} \right)\;\;\;,\;\;\; k=\frac{2\pi m}{N_s}\;\; (m=1,..., N_s-1)
\end{equation}
is the Bogoliubov Hamiltonian describing excitations above the condensate. The single-particle dispersion entering the Bogoliubov spectrum is given by
\begin{equation}
\xi_k(t) = \epsilon_k(t) - \epsilon_0 + U N/N_s,
\end{equation}
where $N$ is the total particle number, $\epsilon_k(t) = -2J \cos\left[k + \phi(t)/N_s\right]$, and $\epsilon_0$ denotes the zero-momentum single-particle energy prior to the onset of the driving.

The time-dependent supercurrent density is given by
\begin{equation}\label{eqn:i-bogoliubov-1}
I(t) = 2\omega_J\sin\left[\frac{\phi(t)}{N_s}\right] n_0(t),
\end{equation}
where $\omega_J=J/\hbar$ and $n_0(t)$ denotes the time-dependent condensate density. Conservation of the total particle number $N$ implies
\begin{equation}
n_0(t) + \sum_{k\neq 0} n_k(t) = \frac{N}{N_s},
\end{equation}
where $n_{k\neq 0}(t)$ represents the particle density in the excited momentum states. Moreover, assuming that all particles initially occupy the zero-momentum condensate state leads to the initial condition $n_k(0) = \delta_{k,0},N/N_s$. Using these relations, Eq.~(\ref{eqn:i-bogoliubov-1}) can be recast as
\begin{equation}\label{eqn:i-mf}
I(t) = \frac{2\omega_J N}{N_s} \sin\left[\frac{\phi(t)}{N_s}\right]
\left( 1 - \frac{1}{N} \sum_{k\neq 0} N_k(t) \right),
\end{equation}
where $N_k(t)$ denotes the time-dependent occupation of the momentum state $k$. 
In the following, we employ the non-equilibrium Green’s function formalism to compute $N_k(t)$.

Using Eq.~(\ref{eqn:h-bogoliubov}) we find the equation of motion for the field operator $\hat{\Psi}_k=(\bk{k},\bdk{-k})^T$,
\begin{equation}
    i\hbar\frac{\partial\hat{\Psi}_k}{\partial t}=\mathcal{H}_k(t)\hat{\Psi}_k,
\end{equation}
where,
\begin{equation}
    \mathcal{H}_k(t)= \begin{pmatrix}
\xi_k(t) & U\frac{N}{N_s} \\
-U\frac{N}{N_s} & \;\;-\xi_{-k}(t)
\end{pmatrix}
\end{equation}

We can use the field operators to respectively build the retarded, and lesser Green function (GF) matrices,
\begin{equation}
    \check{G}^R(k,t,t')=-i\Theta(t-t')\langle[\hat{\Psi}_k(t),\hat{\Psi}_k^\dagger (t')]_\otimes\rangle\;\;,\;\;\check{G}^<(k,t,t')=-i\langle\hat{\Psi}_k^\dagger (t')\otimes\hat{\Psi}_k(t)\rangle,
\end{equation}
where the commutator is taken with respect to the tensor product, i.e. $[A,B]_\otimes = A\otimes B - B\otimes A$. 

The equation of motion for the retarded GF at $t'=0$ is found to be,
\begin{equation}
    \frac{\partial \check{G}^R(k,t,0)}{\partial t}=-i\delta(t)\sigma_z -\frac{i}{\hbar}\mathcal{H}_k \check{G}^R(k,t,0),
\end{equation}
whose formal solution is,
\begin{equation}
    \check{G}^R(k,t,0)=-i\Theta(t)\hat{\mathcal{T}}\left\{ e^{-\frac{i}{\hbar}\int_{0}^t \mathcal{H}_k(s) ds}\right\}\sigma_z,
\end{equation}
where $\hat{\mathcal{T}}$ is the time-ordering operator, ans $\sigma_z$ is the $z$-Pauli matrix. To obtain an approximate analytical expression for the retarded GF, one can use Magnus expansion,
\begin{equation}
    \hat{\mathcal{T}}\left\{ e^{-\frac{i}{\hbar}\int_{0}^t \mathcal{H}_k(s) ds}\right\}\approx \exp\left\{-\frac{i}{\hbar}\int_{0}^t ds_1\mathcal{H}_k(s_1)-\frac{1}{2\hbar^2}\int_{0}^t ds_1\int_{0}^{s_1}ds_2[\mathcal{H}_k(s_1),\mathcal{H}_k(s_2)]+...\right\}.
\end{equation}
Keeping only the first term in the expansion and after some algebraic manipulations we obtain,
\begin{equation}\label{eqn:gr-approx}
    \check{G}^R(k,t,0)\approx -i \Theta(t)e^{i\omega_k t}\left[\cos(\gamma_k t)\sigma_0+\frac{1}{\gamma_k t}\sin(\gamma_k t)M\right],
\end{equation}
where $\sigma_0$ is the identity matrix, $\gamma_k=\sqrt{(\Omega_k+\omega_U)^2-\omega_U^2}$, and
\begin{equation}
    M=-i(\Omega_k+\omega_U)t\sigma_z+(\omega_U t)\sigma_y
\end{equation}
The frequency associated to the interaction energy is $\omega_U=U N/(\hbar N_s)$, while,
\begin{equation}
    \omega_k=2\omega_J\sin(k)\sum_{n=-\infty}^\infty (-1)^n J_n(\tilde{A})\left[\frac{\cos(\Delta\omega_n t-n\theta)-\cos(n\theta)}{\Delta\omega_n t}\right]
\end{equation},
\begin{equation}
    \Omega_k=2\omega_J\left\{1-\cos(k)\sum_{n=-\infty}^\infty (-1)^n J_n(\tilde{A})\left[\frac{\sin(\Delta\omega_n t-n\theta)+\sin(n\theta)}{\Delta\omega_n t}\right]\right\},
\end{equation}
and $\omega_J=J/\hbar$.

We cant then use the retarded GF matrix to calculate the lesser GF from the relation,
\begin{eqnarray}\label{eqn:glesser}
    \check{G}^<(k,t,t')&=&\check{G}^R(k,t,0)\sigma_z\check{G}^<(k,0,0)\sigma_z\check{G}^A(k,0,t')\nonumber\\
    &=&\check{G}^R(k,t,0)\sigma_z\check{G}^<(k,0,0)\sigma_z\left[\check{G}^R(k,0,t')\right]^\dagger,
\end{eqnarray}
with the initial condition,
\begin{equation}
\check{G}^<(k,0,0)=-i\begin{pmatrix}
    \langle \bdk{k}(0)\bk{k}(0)\rangle  &  \langle \bk{k}(0)\bk{k}(0)\rangle\\
    \langle \bdk{k}(0)\bdk{k}(0)\rangle &  \langle \bk{k}(0)\bdk{k}(0)\rangle
\end{pmatrix}
=
-i\begin{pmatrix}
    N\delta_{k,0} & N\delta_{k,0}\\
    N\delta_{k,0} & \;\;1+N\delta_{k,0}
\end{pmatrix}.    
\end{equation}
The particle number in a state with momentum $k$ can then be obtained as,
\begin{equation}\label{nk-lesser}
    N_k(t)=\langle \bdk{k}(t)\bk{k}(t)\rangle = i \check{G}_{11}^<(k,t,t),
\end{equation}
which after using Eqs.~(\ref{eqn:gr-approx}) and (\ref{eqn:glesser}) reduces to,
\begin{equation}\label{eqn:nk}
    N_{k\neq 0}(t)\approx \Theta(t) \frac{\omega_U^2}{2\Omega_k(\Omega_k+2\omega_U)}\left[1-\cos(2\gamma_k t)\right].
\end{equation}
Finally, the current can be computed by substituting Eq.~(\ref{eqn:nk}) into Eq.~(\ref{eqn:i-mf}).

In the near-resonant regime, one can neglect the rapid time oscillations of the contributions to $\Omega_k$ when evaluating the time-averaged current. Consequently, in the vicinity of the $m$th resonance ($\Delta\omega_m=\omega_B-m\omega\approx 0$), $\Omega_k$ and $\gamma_k$ can be replaced by their respective approximations,
\begin{equation}
    \Omega_{km}=2\omega_J\left[1-(-1)^m J_m(\tilde{A})\cos(k)\cos(m\theta)\right]\;\;,\;\;\gamma_{km}=\sqrt{(\Omega_{km}+\omega_U)^2-\omega_U^2}.
\end{equation}
Within this approximation, the total time-averaged supercurrent near the $m$th resonance becomes,
\begin{eqnarray}\label{eqn:i-ave-cond}
    & &\frac{\langle I\rangle_{\tau}^{U}}{I_0}=\frac{\langle I\rangle_{\tau}}{I_0}\left(N-\frac{1}{2}\sum_{k\neq 0}\frac{\omega_U^2}{\gamma_{km}^2}\right)\\
    &+&\frac{(-1)^m}{4}J_m(\tilde{A})\sum_{k\neq 0}\sum_{s=\pm1}\left\{\frac{\omega_U^2}{\gamma_{km}^2} {\rm sinc}\left[\frac{(\Delta\omega_m+2s\gamma_{km})\tau}{2}\right]\sin\left[\frac{(\Delta\omega_m+2s\gamma_{km})\tau}{2}-m\theta\right]\right\}.\nonumber
\end{eqnarray}
is the time-averaged supercurrent carried by the condensate.
Notably, in the limit $\omega_U\rightarrow 0$ (i.e., $U\rightarrow 0$), Eqs.~(\ref{eqn:i-ave-cond})
reduces to the time-averaged current of the non-interacting case [see Eq.~(\ref{corriente promedio pre sinc}) in the main text].

\bibliographystyle{apsrev4-2}
\bibliography{bibliografia-inercia}

\end{document}